# Butterfly-like anisotropic magnetoresistance and angle-dependent Berry phase in Type-II Weyl semimetal WP$_2$ *

Kaixuan Zhang(张凯旋)[1†], Yongping Du(杜永平)[2†], Pengdong Wang(王鹏栋)[3], Laiming Wei(魏来明)[1], Lin Li(李林)[1], Qiang Zhang(张强)[1], Wei Qin(秦维)[1], Zhiyong Lin(林志勇)[1], Bin Cheng(程斌)[1], Yifan Wang(汪逸凡)[1], Han Xu(徐晗)[1], Xiaodong Fan(范晓东)[1], Zhe Sun(孙喆)[3,5], Xiangang Wan(万贤纲)[4,5**], and Changgan Zeng(曾长淦)[1**]

[1] International Center for Quantum Design of Functional Materials, Hefei National Laboratory for Physical Sciences at the Microscale, CAS Key Laboratory of Strongly Coupled Quantum Matter Physics, Department of Physics, and Synergetic Innovation Center of Quantum Information & Quantum Physics, University of Science and Technology of China, Hefei, Anhui 230026, China

[2] Department of Applied Physics and Institution of Energy and Microstructure, Nanjing University of Science and Technology, Nanjing, Jiangsu 210094, China

[3] National Synchrotron Radiation Laboratory, University of Science and Technology of China, Hefei, Anhui 230029, China

[4] National Laboratory of Solid State Microstructures and Department of Physics, Nanjing University, Nanjing, Jiangsu 210093, China

[5] Collaborative Innovation Center of Advanced Microstructures, Nanjing University, Nanjing, Jiangsu 210093, China

*Supported by the National Natural Science Foundation of China under Grant Nos. 11974324, 11804326, U1832151, and 11674296, Strategic Priority Research Program of Chinese Academy of Sciences under Grant No. XDC07010000, National Key Research and Development Program of China under Grant No. 2017YFA0403600, Anhui Initiative in Quantum Information Technologies under Grant No. AHY170000, Hefei Science Center CAS under Grant No. 2018HSC-UE014, Jiangsu Province Science Foundation for Youth under Grant No. BK20170821, National Science Foundation of China for Youth under Grant No. 11804160, and Anhui Provincial Natural Science Foundation under Grant No. 1708085MF136. A portion of this work was performed on the Steady High Magnetic Field Facilities, High Magnetic Field Laboratory, CAS. We thank Li Pi and Chuanying Xi for their experimental support and helpful discussions.

[†]K. Zhang, and Y. Du contributed equally to this work.

[**]Corresponding authors. Email: xgwan@nju.edu.cn; cgzeng@ustc.edu.cn

**Abstract** Weyl semimetal emerges as a new topologically nontrivial phase of matter, hosting low-energy excitations of massless Weyl fermions. Here, we present a comprehensive study of the type-

II Weyl semimetal WP$_2$. Transport studies show a butterfly-like magnetoresistance at low temperature, reflecting the anisotropy of the electron Fermi surfaces. The four-lobed feature gradually evolves into a two-lobed one upon increasing temperature, mainly due to the reduced relative contribution of electron Fermi surfaces compared to hole Fermi surfaces for the magnetoresistance. Moreover, angle-dependent Berry phase is further discovered from the quantum oscillations, which is ascribed to the effective manipulation of the extremal Fermi orbits by the magnetic field to feel the nearby topological singularities in the momentum space. The revealed topological characters and anisotropic Fermi surfaces of WP$_2$ substantially enrich the physical properties of Weyl semimetals and hold great promises in topological electronic and Fermitronic device applications.

**PACS:** 03.65.Vf, 71.20.Gj, 72.15.-v, 75.47.-m

---

Type-II Weyl semimetals possess highly anisotropic electronic structures, e.g., significantly tilted Weyl cones and anisotropic Fermi surfaces.[1,2] Such anisotropies can be exploited to produce a wide spectrum of exciting physical properties, including the anisotropic version of chiral anomaly,[3,4] photocurrent response,[5,6] magnetoresistance (MR),[7] and plasma mirror behavior,[8,9] etc. Recently, WP$_2$ was revealed to be a new robust Type-II Weyl semimetal with both anisotropic hole and electron Fermi surfaces.[10-13] Since MR behavior is intimately related to the topology of the Fermi surface,[14] the anisotropies of both hole and electron Fermi surfaces may collectively lead to exotic anisotropic magnetoresistance (AMR) patterns.

In general, degenerate points (such as band crossings) may exist in the momentum space of a material, which act as monopoles to create Berry fields.[15] If an electron is moving in a closed orbit near a degenerate point in the momentum space, it acquires an additional nonzero Berry phase ($\phi_B$) as the integral of the Berry field flux, i.e., Berry curvature, from the nearby monopoles.[16-20] The nonzero Berry phase has been revealed in graphene,[21,22] topological insulators,[23,24] Rashba semiconductors,[20] Dirac semimetals[25] and Weyl semimetals,[26] manifested in the quantum transport effects, in particular the Shubnikov-de Haas (SdH) quantum oscillations.[16,20-27] For a Weyl semimetal without inversion symmetry, in addition to the Weyl nodes (band crossing points of the valence and conductance bands), other degenerate points may also present due to the strong spin-orbit coupling. Such topological features further enrich the fascinating physics of Weyl semimetals, and render more possibilities in designing future topological electronic devices.

In the present work, we investigate the Type-II Weyl semimetal $WP_2$ by electronic transport. A four-lobed butterfly-like AMR emerges at low temperatures and gradually develops into a two-lobed one on increasing temperature, mainly owing to the anisotropic electron Fermi surfaces. Moreover, angle-dependent Berry phase is further discovered from the quantum oscillations, which is ascribed to the topological singularities in the momentum space.

High quality $WP_2$ single crystals were grown by chemical vapor transport method using iodine as the transport agent.[28] The P, $WO_3$ and $I_2$ sources were mixed and sealed in an evacuated quartz tube, and the $WP_2$ crystals were grown in a two-zone furnace with a temperature gradient of 1000 ˚C (source) to 900 ˚C (sink) for 10 days. The stoichiometric ratio for W:P was confirmed to be 1:2 by energy-dispersive X-ray spectroscopy. Low-field transport measurements were performed on $WP_2$ crystals using the Resistivity with rotator option in a Quantum Design physical property measurement system with the highest magnetic field up to 14 T. High-field transport measurements were carried out using standard ac lock-in techniques with a He-3 cryostat and a dc-resistive magnet (~ 35 T) at the High Magnetic Field Laboratory of Chinese Academy of Sciences (CAS). In order to improve the electrical contact, Au/Ti electrodes with thickness of 75 nm/5 nm were first deposited on the sample by hard-mask method, and then gold wires were used to make contacts between the chip carrier and the Au/Ti electrodes with silver paint.

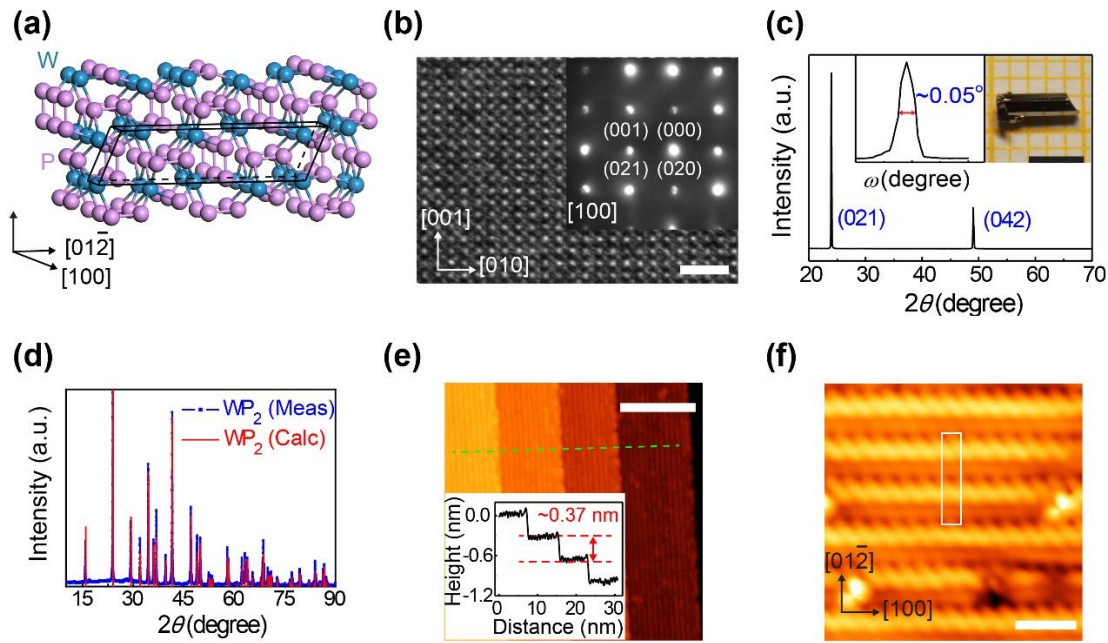

**Fig. 1.** Structure characterizations. (a) Crystallographic structure of $WP_2$ with the (021) surface. The blue and purple balls represent W and P atoms, respectively. The black lines indicate a unit cell. (b) TEM and SAED of a $WP_2$ single crystal. The scale bar is 2 nm. (c) Single-crystal XRD pattern of a $WP_2$ single crystal. The insets show the FWHM of the (021) peak in the rocking curve (~ 0.05°) and the optical image of a shiny as-grown crystal. The scale bar is 3 mm. (d) Powder XRD pattern of grinded $WP_2$ crystals. The measured peaks (blue) perfectly coincide with the calculated results (red). (e) STM image of the cleaved (021) surface with well-resolved terraces. The inset shows the height profile along the green dashed line. The scale bar is 10 nm. (f) STM image of the cleaved (021) surface with atomic resolution. The white solid rectangle denotes a surface unit cell. The scale bar is 1 nm.

As depicted in Fig. 1(a), the grown $WP_2$ crystals possess an orthorhombic structure (β-phase).[29,30] Figure 1(b) shows the high-resolution transmission electron microscopy (TEM) image and the corresponding selected area electron diffraction (SAED) pattern, which demonstrate the high quality of the single crystal at atomic scale. The crystalline long axis, i.e., the growth direction is the [100] direction. The biggest crystal face of the as-grown $WP_2$ single crystals is the (021) plane (up to 3 mm × 2 mm), confirmed by X-ray diffraction (XRD) measurement (Fig. 1(c)) with a small full width at half maximum (FWHM) of the rocking curve (~ 0.05°). The powder XRD results after single crystals grinded into powders are displayed in Fig. 1(d), and the perfect match of the experimental result and calculated pattern for β-phase WP₂ further validates the high crystalline quality and purity. Figure 1(e) shows the typical STM image of the cleaved (021) surface with terrace height ~ 0.37 nm, very close to the (021) layer spacing ~ 0.371 nm.[30] The high-resolution STM image in Fig. 1(f) shows atomic chain structure along the [100] direction, and the white rectangle region represents a unit cell of the (021) surface.

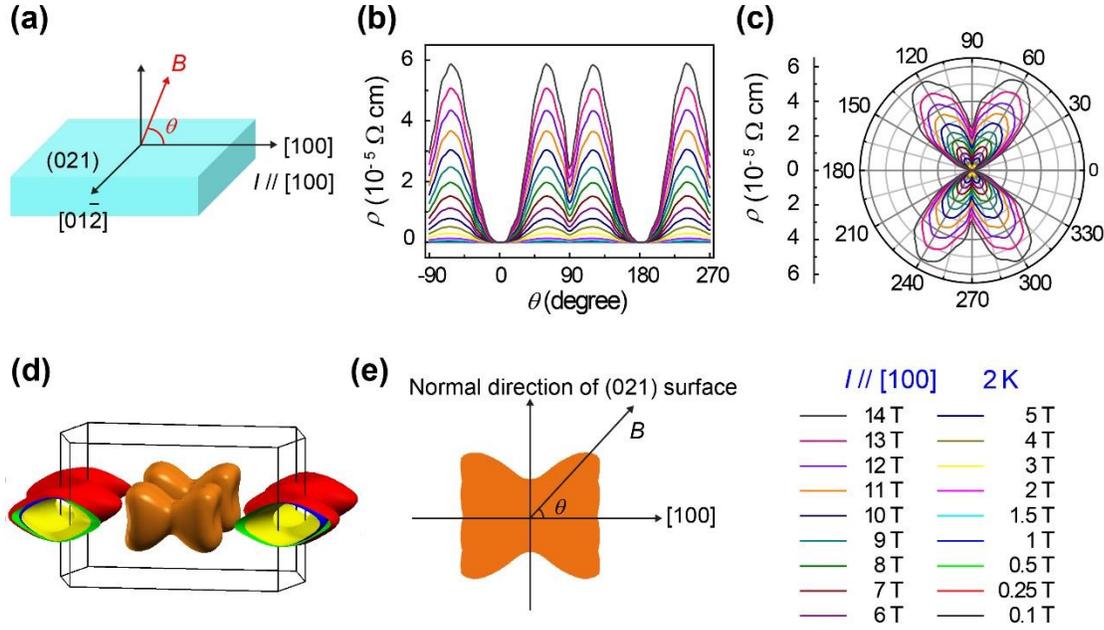

**Fig. 2.** Butterfly-like AMR. (a) Schematic of the measurement configuration. The current $I$ is injected along the [100] direction. The magnetic field $B$ is rotated in the plane consisting of the [100] direction and the normal direction of the (021) surface. The angle between the magnetic field and the [100] direction is denoted as $\theta$. (b) Resistivity $\rho$ as a function of $\theta$ at 2 K under various magnetic fields. (c) Polar plot of $\rho$ as a function of $\theta$, illustrating the butterfly-like AMR with four lobes. (d) Calculated Fermi surfaces of WP₂. The bow-tie-like closed pockets are electron Fermi surfaces, while the spaghetti-like open pockets are hole Fermi surfaces. (e) Projected electron Fermi surfaces in the rotation plane of $B$.

The basic transport properties including the temperature and magnetic field dependences of the WP₂ resistivity are displayed in Fig. S1 in the supplementary material. Here we focus on the interesting AMR behavior of WP₂. As illustrated in Fig. 2(a), the magnetic field ($B$) was rotated in the plane consisting of the [100] direction and the normal direction of the (021) surface, and the current ($I$) was injected along the [100] direction. The angle between $B$ and the [100] direction is denoted as $\theta$. As shown in Figs. 2(b,c), when $B$ is rotated, the resistivity ($\rho$) peaks at $\theta = 60°$, $120°$, $240°$, and $300°$, forming a butterfly-like pattern with four lobes. This is in sharp contrast to the two-lobed AMR reported previously for WP$_2$ with $B$ rotating in the [100]-[010] or [010]-[001] planes.[11,31,32]

Since the electronic transport is dominated by the carriers in the vicinity of the Fermi level,[33] the MR behavior is intimately related to the topology of the Fermi surface.[14] Theoretical calculations reveal that the Fermi surfaces of WP$_2$ consist of two pairs of electron and hole pockets which are split by the spin-orbit coupling,[9-11,31,32] as depicted in Fig. 2(d). The electron Fermi surfaces are closed with a bow-tie-like shape, whereas the hole Fermi surfaces are open with a spaghetti-like shape extending along the [010] direction.[11,31,32] It is noted that the projected electron Fermi surfaces on the rotation plane of $B$ are also bow-tie-like as shown in Fig. 2(e), similar to the AMR shape. As detailed in the supplementary material, the butterfly-like AMR is largely determined by the intrinsic anisotropy of the electron Fermi surfaces while both the anisotropic hole and electron Fermi surfaces collectively contribute to the AMR.

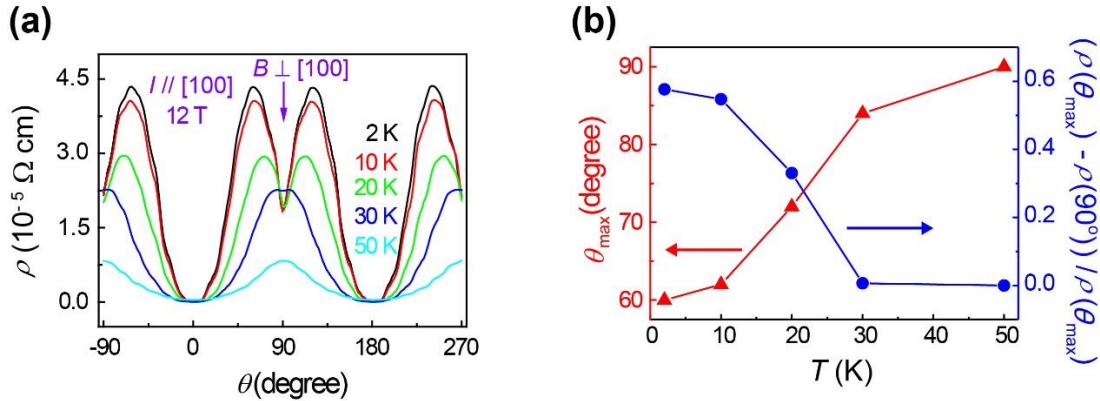

**Fig. 3.** Temperature-dependent evolution of the AMR. (a) $\rho$ as a function of $\theta$ with $B = 12$ T at various temperatures. (b) The angle of the resistivity peak ($\theta_{max}$) and the AMR ratio $[\rho(\theta_{max}) - \rho(90°)]/\rho(\theta_{max})$ as functions of $T$.

Angle-dependent magnetoresistance measurements were systemically performed at various magnetic fields and temperatures, and the results are displayed in Figs. 3(a) and S2. It is evident that when the temperature rises, the butterfly-like AMR with four lobes gradually evolves into a two-lobed AMR. To quantitatively characterize this feature, the angle where the resistivity peaks ($\theta_{max}$) and the AMR ratio defined as $[\rho(\theta_{max}) - \rho(90°)]/\rho(\theta_{max})$ are plotted as functions of temperature, as shown in Fig. 3(b). It can be seen that $\theta_{max}$ varies from 60° to 90° with increasing temperature, whereas the AMR ratio vanishes. Such AMR evolution upon increasing temperature is mainly due to the reduced relative contribution of electron Fermi surfaces compared to hole Fermi surfaces for the magnetoresistance.

Exploration of the exotic effects originated from the particularity, such as strong anisotropy, of the Fermi surfaces, i.e., Fermitronics, has emerged in the field of electronics.[34] Here we demonstrate that the AMR effect in WP$_2$ can be effectively engineered by selectively tailoring the anisotropy of the carrier trajectories at the three-dimensional Fermi surfaces by, for example, varying the direction of the applied magnetic field and temperature.

The nontrivial topological nature is the most attractive property of the Weyl semimetals, and the Berry phase can be experimentally determined by the phase information of the SdH quantum oscillations.[18,22-30] Next we exploit the Berry phase of WP$_2$ using the quantum oscillation as an effective probe. Figure 4(a) shows the resistivity ($\rho$) as a function of $B$ at various $\theta$s, with $B$ up to 33 T. The magnetoresistance oscillations appear at high fields, and are attributed to the SdH quantum oscillations.[20-27] The SdH oscillations originate from the quantization of the closed electronic orbits under applied magnetic field, following the Lifshitz-Onsager quantization rule[35-37] incorporating the Berry phase.[15,20] To better illustrate the SdH quantum oscillations, the classical nonoscillatory background is subtracted from $\rho$[11,31,32,34] (see more details in the supplementary material), and the oscillatory component ($\Delta\rho$) is depicted in Fig. 4(b).

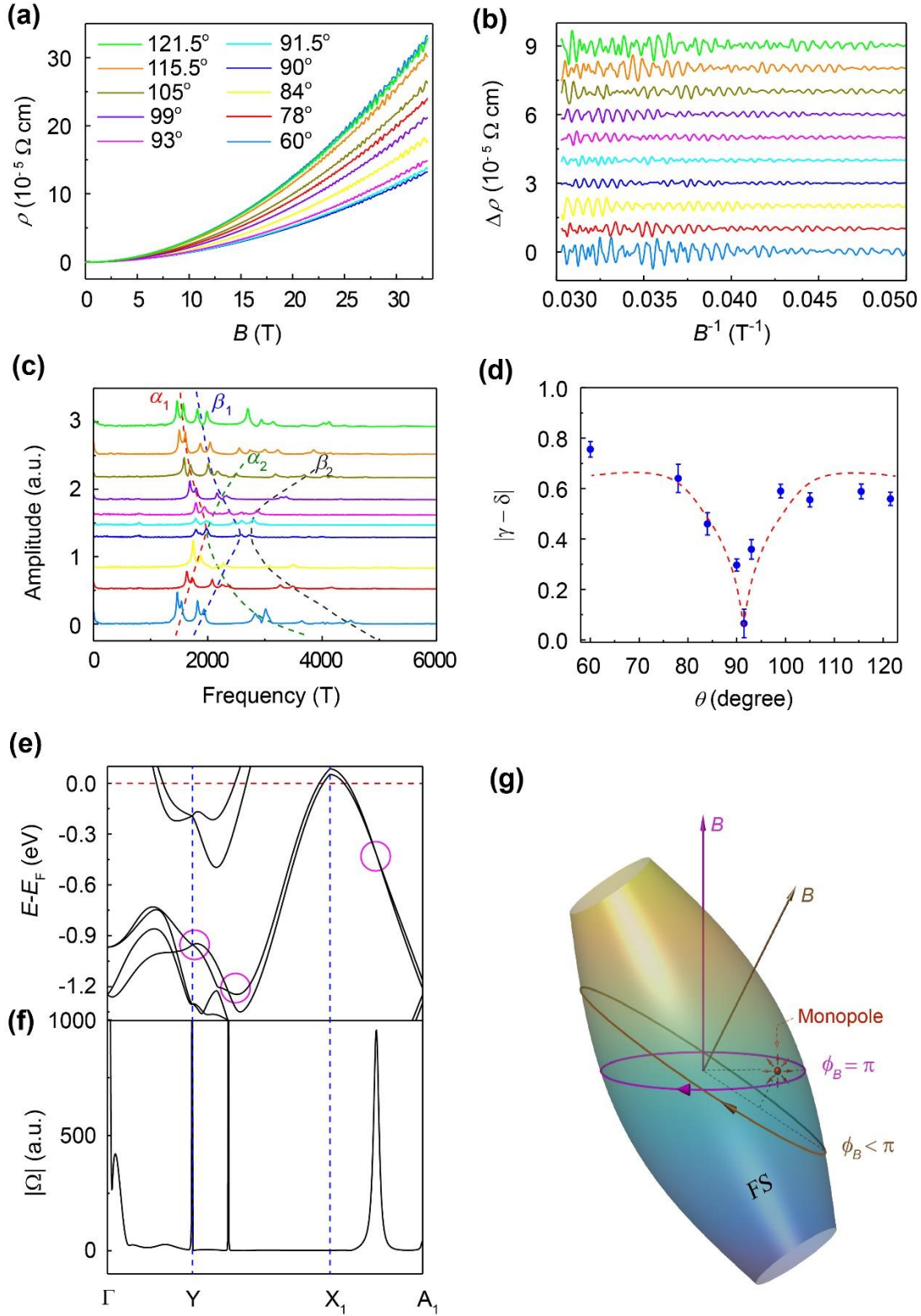

**Fig. 4.** Quantum oscillations. (a) Resistivity $\rho$ as a function of $B$ at 2 K for various $\theta$s. (b) The oscillatory component ($\Delta\rho$) as a function of $1/B$ after subtraction of a smooth MR background from $\rho$. (c) FFT spectra of $\Delta\rho$ at various $\theta$s. The dashed curves are the calculated angular dependences of the SdH frequencies associated with the individual extremal cross-sectional orbits of the hole Fermi surface pockets ($\alpha_1$, $\alpha_2$, $\beta_1$, and $\beta_2$). (d)

Magnitude of the total phase $|\gamma - \delta|$ as a function of $\theta$ for the $\alpha_1$ branch of the hole Fermi surfaces. The red dashed curve is guide to the eyes for the evolution trend of the total phase. (e) Calculated bulk band structures of WP$_2$ along the $\Gamma$-Y-X$_1$-A$_1$ direction. The pink circles mark the band crossing points of the two valence subbands. (f) Theoretically calculated magnitude of the Berry curvature $|\Omega|$ of the valence bands along the $\Gamma$-Y-X$_1$-A$_1$ direction. $|\Omega|$ peaks at the band crossing points. (g) Schematic of the angle-dependent Berry phase in the presence of a monopole when rotating the magnetic field.

Figure 4(c) displays the fast Fourier transform (FFT) spectra of $\Delta\rho$ at different $\theta$s. The multiple peaks in the FFT spectra can be ascribed to different extremal cross-sectional orbits of individual hole Fermi surface pockets by comparing their frequencies with the theoretically calculated values,[31] as justified in the supplementary material. The frequencies of these peaks vary with $\theta$, indicating the quasi-two-dimensional nature of the Fermi surface pockets.[31,32,34] It is noted that the FFT peak associated with the $\alpha_1$ branch of the hole Fermi surfaces splits into double peaks, which may originate from the fine structures beyond the calculation resolution.[38] On the other hand, no pronounced FFT peak related to the electron Fermi surfaces is observed (see more details in the supplementary material).

Then we focus on the Berry phase for the $\alpha_1$ branch of the hole Fermi surfaces, because its FFT peaks are well resolved at all measured angles. Following the methods described in previous studies,[26,39,40] the SdH oscillations components associated with the double FFT peaks of the $\alpha_1$ branch are extracted, and the corresponding total phase magnitude $|\gamma - \delta|$ is derived from the two-band Lifshitz-Kosevich fitting (see Figs. 4(d) and S3, and more details in the supplementary material). Here $\gamma = 1/2 - \phi_B/2\pi$ is the Onsager phase and $\delta$ the phase shift equals to zero for the two-dimensional (2D) or $\pm 1/8$ for the three-dimensional (3D) cases.[26,39,40] It can be seen that $|\gamma - \delta|$ drops from ~0.8 at 60° to ~0.07 at 91.5° and then increases back to ~0.6 near 120°, indicating a highly tunable Berry phase. In particular, $|\gamma - \delta|$ is almost zero at 91.5°, suggesting a Berry phase close to $\pi$ (A $\pi$ Berry phase lead to $|\gamma - \delta| = 0$ (2D) or $|\gamma - \delta| = 1/8$ (3D)[26,39,40]).

Similar angle-dependent Berry phase was also reported in ZrSiS[34] recently, whereas the underlying mechanism is unclear. In the present study, such behavior can be attributed to the shape of the extremal Fermi orbit perpendicular to the applied magnetic field $B$ and its relative position with respect to the Berry curvature monopoles. Based on our theoretical calculations, there are quite a few crossing points for the two valence subbands of $WP_2$, and some of them in the high-symmetry line are depicted in Fig. 4(e). As shown in Fig. 4(f), the Berry curvature sharply peaks at these band crossing points, which can be theoretically regarded as monopoles, serving as the source or sink of the Berry field or Berry curvature in the momentum space.[15,17] Although the Berry curvature is generated by these monopoles located at several hundreds of meV below the Fermi level as marked in Fig. 4(e), the corresponding Berry connection tends to extend through the whole energy bands.[27] In the presence of an external magnetic field, carriers near the Fermi surface move along the extremal cross-sectional orbit, accumulating a Berry phase equals to the path integral of the Berry connection.[16-20] When the magnetic field is rotated, the shape of extremal cross-sectional orbit and its relative position with respect to these monopoles vary accordingly, thereby altering the path integral of Berry connection along the orbit, namely, the Berry phase. Figure 4(g) depicts a simplified case with only a single monopole: When the extremal cross-sectional orbit of carriers perpendicular to $B$ happens to enclose the monopole, the corresponding Berry phase equals to $\pi$.[16,17] While $B$ is rotated to make the extremal cross-sectional orbit away from the monopole, the Berry phase decrease gradually.

It is well established that the nonzero Berry phase of Fermi surfaces could induce exotic effects, e.g., half-integer quantum Hall effect[21,22] and valley Hall effect[41] in 2D graphene. In contrast to the 2D systems, the Berry phase for carriers moving in varying closed orbits in 3D systems can be effectively tuned as demonstrated in the present study. Such high tunability of Berry phase in 3D systems extends a new dimension to control the intriguing topological quantum effects.

In summary, electronic transport reveals a butterfly-like AMR at low temperatures and its temperature-dependent evolution, mainly arising from the anisotropic electron Fermi surfaces.

Moreover, an angle-dependent Berry phase is demonstrated from the SdH quantum oscillations, which can be attributed to the topological singularities in the momentum space. Our findings not only inspire deep insights into the understanding of the topological and Fermi surface properties of Weyl semimetals, but also promise potential applications in corresponding topological electronic and Fermitronic devices.

**Supplementary Material for:**

# Butterfly-like anisotropic magnetoresistance and angle-dependent Berry phase in Type-II Weyl semimetal WP$_2$ *

Kaixuan Zhang(张凯旋)[1†], Yongping Du(杜永平)[2†], Pengdong Wang(王鹏栋)[3], Laiming Wei(魏来明)[1], Lin Li(李林)[1], Qiang Zhang(张强)[1], Wei Qin(秦维)[1], Zhiyong Lin(林志勇)[1], Bin Cheng(程斌)[1], Yifan Wang(汪逸凡)[1], Han Xu(徐晗)[1], Xiaodong Fan(范晓东)[1], Zhe Sun(孙喆)[3,5], Xiangang Wan(万贤纲)[4,5**], and Changgan Zeng(曾长淦)[1**]

[1] International Center for Quantum Design of Functional Materials, Hefei National Laboratory for Physical Sciences at the Microscale, CAS Key Laboratory of Strongly Coupled Quantum Matter Physics, Department of Physics, and Synergetic Innovation Center of Quantum Information & Quantum Physics, University of Science and Technology of China, Hefei, Anhui 230026, China

[2] Department of Applied Physics and Institution of Energy and Microstructure, Nanjing University of Science and Technology, Nanjing, Jiangsu 210094, China

[3] National Synchrotron Radiation Laboratory, University of Science and Technology of China, Hefei, Anhui 230029, China

[4] National Laboratory of Solid State Microstructures and Department of Physics, Nanjing University, Nanjing, Jiangsu 210093, China

[5] Collaborative Innovation Center of Advanced Microstructures, Nanjing University, Nanjing, Jiangsu 210093, China

[†] K. Zhang, and Y. Du contributed equally to this work.

[**] Corresponding authors. Email: xgwan@nju.edu.cn; cgzeng@ustc.edu.cn

# Supplementary notes

1. **Electronic band structure calculations.**

The electronic band structure calculations were carried out using the full potential linearized augmented plane-wave method as implemented in the WIEN2K package.[1] For topological material (TM), the band order is a key to determine the topological trivial or non-trivial bands.[2] The modified Becke-Johnson (MBJ) potential[2,3] had been used to successfully predict a wide variety of TMs.[4,5] Here, we apply the MBJ potential together with the local-density approximation for the correlation potential (MBJLDA) to get the accurate band order and the band inversion strength.[3] The plane-wave cutoff parameter $R_{MT}K_{max}$ was set to be 7 and a 24× 24× 15 mesh was used for the BZ integral. The spin-orbit coupling was treated using the second-order variational procedure. The Fermi surfaces were generated via a more refined k-point mesh of 40 × 40 × 24. Fermi surface sheets were visualized using the XCrysden software.[6] The angular dependence of the quantum oscillatory frequencies was calculated through the Skeaf code.[7] The Berry curvature were then calculated within the tight-binding model using the WannierTools software package.[8,9] The hopping parameters were determined from the maximally localized Wannier functions,[10] which were projected from the Bloch state derived from the first-principles calculations.

2. **Temperature and magnetic field dependences of the WP$_2$ resistivity.**

Figure S1(a) displays the resistivity ($\rho$) of WP$_2$ as a function of temperature ($T$) at various magnetic fields ($B$), and Figure S1(b) depicts $\rho$ as a function of $B$ at various temperatures. The magnetic field is perpendicular to the (021) surface. An extremely low residual resistivity (~ 4 n$\Omega$·cm at 0 T and 2 K) and a gigantic magnetoresistance (MR) (~ $10^6$% at 14 T and 2 K) are revealed, comparable to the values reported recently for WP$_2$.[11-13] Such low residual resistivity and gigantic MR can be attributed to the topologically suppressed carrier scattering and perfect electron-hole

compensation in WP$_2$.[13]

3. **Anisotropic magnetoresistance (AMR) arising from the anisotropy of the Fermi surface.**

As depicted in Fig. 2(a) in the main text, the magnetic field $B$ was rotated in the plane consisting of the [100] direction and the normal direction of the (021) surface, and the current was injected along the [100] direction. The angle between $B$ and the [100] direction is denoted as $\theta$. It is noted that the rotation of $B$ is almost in the bisector plane between the [100]-[010] plane and [100]-[001] plane, and thus can be decomposed into a rotation in the [100]-[010] plane and a rotation in the [100]-[001] plane.

For the open hole Fermi surface, when $B$ is rotated in the [100]-[010] plane, the MR is expected to be extremely large along the [010] direction because the anisotropic hole Fermi surface extends along such direction,[11] leading to a strong AMR with two lobes at $\theta = 90°$ and 270°. In contrast, when $B$ is rotated in the [100]-[001] plane, the MR is much weaker and can be neglected because the perpendicular cross-section area of the Fermi surface always remains infinite.[13] Therefore, the MR arising from the anisotropic hole Fermi surface is dominated by the MR in the [100]-[010] plane, forming a global AMR with two lobes and thus cannot account for the observed butterfly-like AMR with four lobes.

On the other hand, two body diagonals of the electron Fermi surfaces are almost in the rotation plane of $B$, therefore the projected electron Fermi surfaces on this plane are also bow-tie-like as shown in Fig. 2(e) in the main text, similar to the AMR shape. When $B$ is perpendicular to the steep regions of the electron FS, the averaged cyclotron mass and cyclotron frequency of carriers travelling in the plane perpendicular to $H$ are expected to be minimal and maximal,[14-17] leading to a large MR. That is, the anisotropy of the electron Fermi surfaces is expected to give rise to a butterfly-like AMR with four MR peaks along the lobes of the electron Fermi surfaces (i.e., $\theta = 45°$, 135°, 225°, and 315°).[14-17] These angles are slightly deviated from the experimental observations, mainly due to the additional contributions from the anisotropic hole Fermi surfaces.

In summary, while both the anisotropic hole and electron Fermi surfaces collectively contribute to the AMR, the observed butterfly-like four-lobe feature of AMR is largely determined by the electron Fermi surface.

4. **Analysis of the Shubnikov-de Haas (SdH) oscillations.**

To better illustrate the SdH quantum oscillations, after fitting $\rho(B)$ with a power law formula $A \times B^n$ (where A is a constant and the power n = 1.8-2.0 for different angles), the classical nonoscillatory background $A \times B^n$ is subtracted from $\rho$,[11-13,18] and the oscillatory component ($\Delta\rho$) is depicted in Fig. 4(b) in the main text.

Table S1 lists the theoretically calculated SdH frequencies corresponding to different extremal cross-sectional orbits of the electron and hole Fermi surfaces at various $\theta$s. Comparing these frequencies with the positions of peaks in the fast Fourier transform (FFT) spectra (Fig. 4(c) in the main text), the multiple FFT peaks can be ascribed to individual Fermi surface pockets. For example, the split double peaks at 1740 and 1870 T for $\theta = 84°$ are attributed to an extremal cross-sectional orbit ($\alpha_1$) of the hole Fermi surface pocket ($\alpha$), whereas the peaks at 2180, 2270, and 3060 T are associated with other extremal cross-sectional orbits ($\alpha_2$, $\beta_1$, and $\beta_2$) of the hole Fermi surface pockets ($\alpha$ and $\beta$). The $\alpha_1$ peak also splits into double peaks at other $\theta$s. Similar splitting was also observed previously in WP$_2$[11] and TaP,[19] and are attributed to the fine structures in the hole pocket beyond the calculation resolution.[19] It is also noted that the calculated SdH frequencies of $\alpha_2$ and $\alpha_1$ branches merge with each other at 90°. Nevertheless, the SdH oscillation component of the $\alpha_1$ branch overwhelms that of the $\alpha_2$ branch at 90°, since the FFT amplitude of the $\alpha_2$ peak is much smaller than that of the $\alpha_1$ peak when $\theta$ is approaching to 90°.

On the other hand, no pronounced peak related to the electron Fermi surfaces is observed in the FFT spectra. Therefore, the SdH oscillations are dominated by the hole Fermi surfaces, whereas the butterfly-like AMR is largely determined by the electron Fermi surfaces as discussed earlier. The quantum mobility may be lower and the corresponding Landau level broadening of electrons may be wider than that of holes, rendering the quantization of the extremal orbits around the electron Fermi surface pockets much more difficult. Hereafter we focus on the Berry phase for the hole branches.

The Berry phase ($\phi_B$) can be extracted from the SdH oscillations by the Lifshitz-Kosevich fitting[18,20,21] or the Landau level index plot.[22] However, the latter is no longer a reliable method to analyze the Berry phase for multiband electronic systems,[20,23] therefore we adopt the former method following previous studies.[20,21]

The SdH oscillation component associated with the $\alpha_1$ branch of the hole Fermi surfaces is extracted using the frequency filtering and inverse FFT method.[18,24,25] The results are shown in Fig. S3. We employ the two-band Lifshitz-Kosevich formula $\Delta\rho \propto e^{\frac{-\pi}{\mu_q B}} \frac{\lambda}{\sinh(\lambda)} \left\{ \cos\left[2\pi\left(\frac{B_{F1}}{B} - \delta + \gamma\right)\right] + \cos\left[2\pi\left(\frac{B_{F2}}{B} - \delta + \gamma\right)\right] \right\}$ to extract the magnitude of the total phase ($|\gamma - \delta| = |1/2 - \phi_B/2\pi - \delta|$) related to the $\alpha_1$ branch.[20] Here $B_{F1}$ and $B_{F2}$ are the SdH frequencies corresponding to the double FFT peaks, $\gamma$ is the Onsager phase, $\delta$ is a phase shift, $\mu_q$ is the hole quantum mobility, $\lambda = \frac{2\pi^2 m^* k_b T}{\hbar e B}$, $m^*$ is the effective mass of the $\alpha_1$ holes, and $k_b$ is the Boltzmann constant. From the temperature dependence of the SdH amplitude (data not shown), $m^*$ is estimated to be 1.4 $m_0$ ($m_0$ is the electron rest mass), comparable to previous results of WP$_2$.[11]

# Supplementary references

**Supplementary figures**

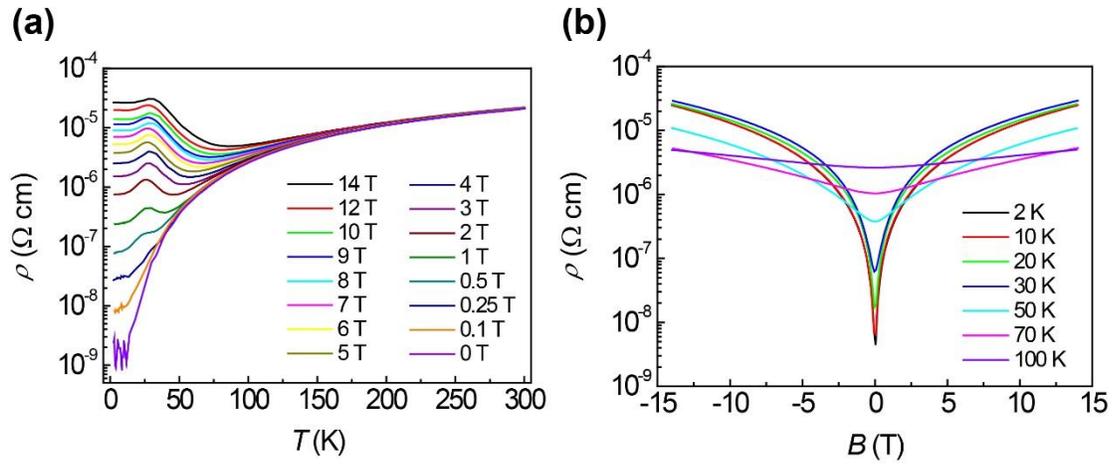

**Fig. S1.** Transport properties of WP$_2$. (a) Resistivity $\rho$ as a function of temperature $T$ at different magnetic fields. (b) $\rho$ as a function of magnetic field $B$ at various temperatures. The magnetic field in (a) and (b) is perpendicular to the (021) plane.

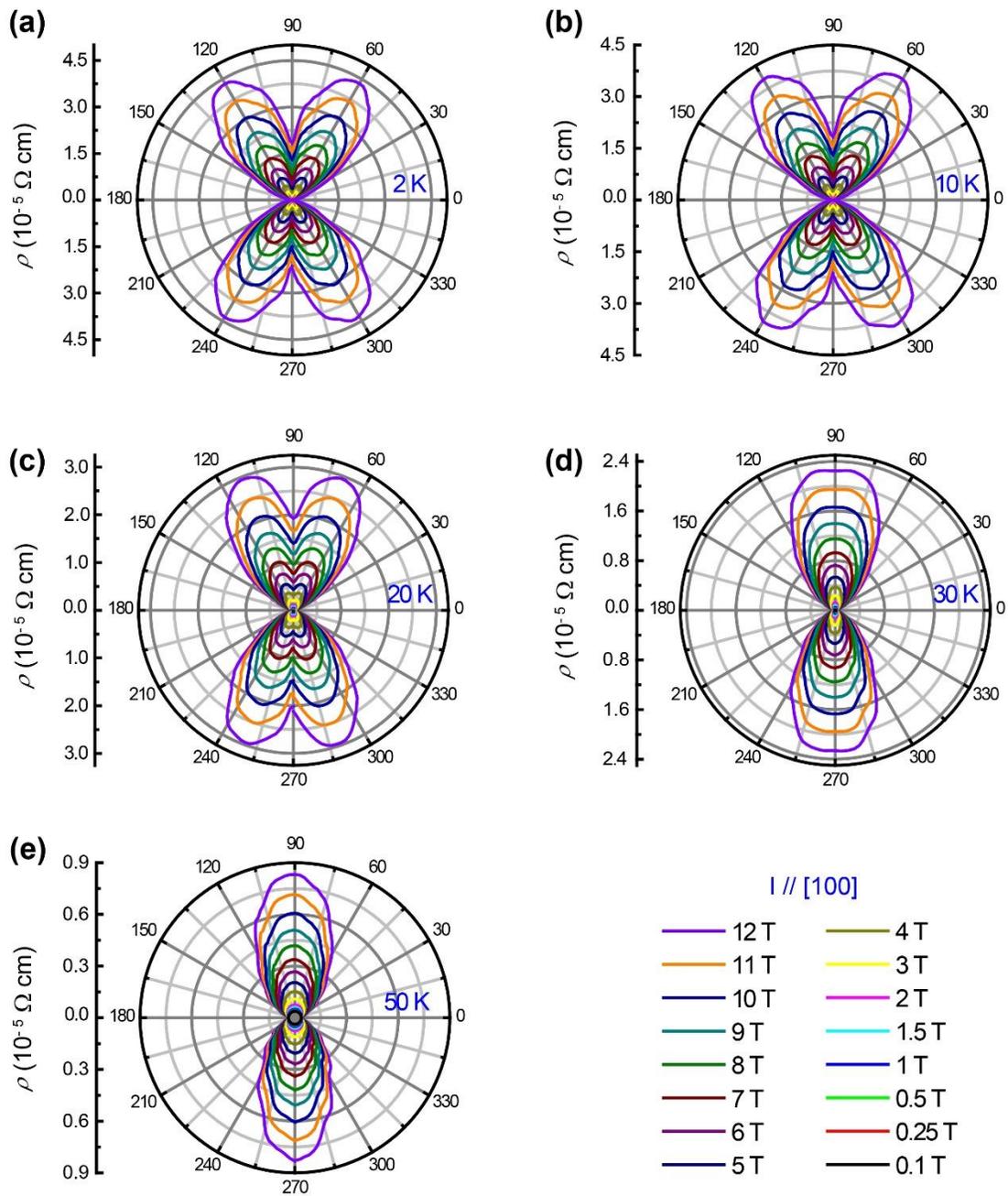

**Fig. S2.** Temperature-dependent evolution of the AMR. (a-e) Polar plots of $\rho$ as a function of $\theta$ with different magnetic fields at 2 K, 10 K, 20 K, 30 K, and 50 K, respectively.

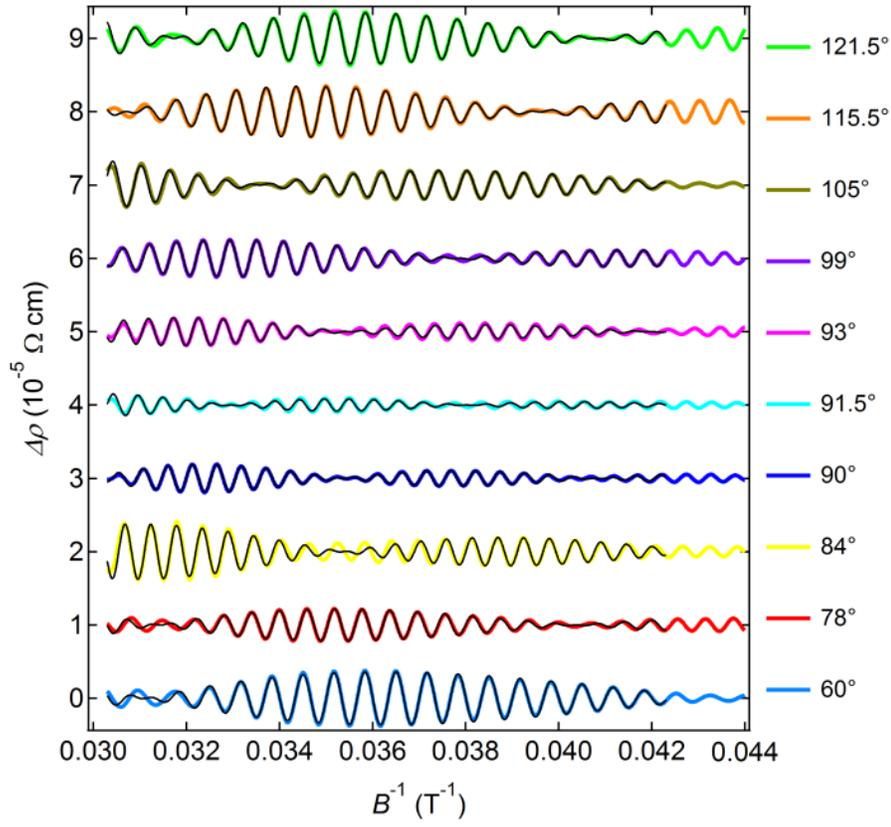

**Fig. S3.** SdH oscillation component of the $\alpha_1$ branch of the hole Fermi surfaces at various $\theta$s. The curves are shifted vertically for clarity. The black curves are the fitting curves based on the Lifshitz-Kosevich model.

# Supplementary table

**Table S1.** Theoretically calculated SdH frequencies associated with individual extremal cross-sectional orbits of the hole ($F_{\alpha 1}$, $F_{\alpha 2}$, $F_{\beta 1}$ and $F_{\beta 2}$) and electron ($F_\delta$ and $F_\gamma$) Fermi surfaces at various $\theta$s.

| Orbits<br>Angles | $\alpha_1$<br>$F_{\alpha 1}$(kT) | $\alpha_2$<br>$F_{\alpha 2}$(kT) | $\beta_1$<br>$F_{\beta 1}$(kT) | $\beta_2$<br>$F_{\beta 2}$(kT) | $\delta$<br>$F_\delta$(kT) | $\gamma$<br>$F_\gamma$(kT) |
|---|---|---|---|---|---|---|
| 121.5° | 1.5864 | 3.1768 | 1.9416 | 4.4489 | 4.5209 | 3.6235 |
| 115.5° | 1.6048 | 2.9061 | 1.9686 | 4.1977 | 4.6627 | 3.6167 |
| 105° | 1.6980 | 2.5373 | 2.1058 | 3.5893 | 4.7736 | 3.6171 |
| 99° | 1.7926 | 2.3161 | 2.2491 | 3.1764 | 4.8249 | 3.6471 |
| 93° | 1.9253 | 2.1016 | 2.4599 | 2.7657 | 4.8583 | 3.6726 |
| 91.5° | 1.9638 | 2.0533 | 2.5262 | 2.7376 | 4.8634 | 3.6772 |
| 90° | 1.9913 | 2.0027 | 2.5966 | 2.7263 | 4.8614 | 3.6736 |
| 84° | 1.8513 | 2.2025 | 2.3424 | 2.9582 | 4.8430 | 3.6590 |
| 78° | 1.7387 | 2.4252 | 2.1680 | 3.3865 | 4.7969 | 3.6276 |
| 60° | 1.5863 | 3.0975 | 1.9428 | 4.3931 | 4.5639 | 3.6227 |